\documentclass[twocolumn,superscriptaddress,aps,showpacs,amsmath,footinbib,bibnotes]{revtex4-1}
\pdfoutput=1
\usepackage{graphicx}
\usepackage{amssymb,amsmath}
\usepackage{textcomp} 
\usepackage[bold]{hhtensor}
\usepackage{natbib}
\usepackage{lipsum}
\usepackage{bm}
\usepackage{hyperref}
\usepackage{multirow}
\usepackage{float}
\usepackage{amsmath}
\usepackage{amssymb}
\usepackage{stmaryrd}
\usepackage{mathrsfs}
\usepackage[dvipsnames]{xcolor}
\hypersetup{
	colorlinks,
	linkcolor={red!80!black},
	citecolor={Blue},
	urlcolor={MidnightBlue}
}

\begin{document}

\author{Boris Desiatov}
\affiliation{John A. Paulson School of Engineering and Applied Sciences, Harvard University, Cambridge, Massachusetts 02138, USA}
\author{Amirhassan Shams-Ansari}
\affiliation{John A. Paulson School of Engineering and Applied Sciences, Harvard University, Cambridge, Massachusetts 02138, USA}
\author{Mian Zhang}
\affiliation{John A. Paulson School of Engineering and Applied Sciences, Harvard University, Cambridge, Massachusetts 02138, USA}
\affiliation{HyperLight Corporation, 501 Massachusetts Ave, Cambridge, MA 02139, USA}
\author{Cheng Wang}
\affiliation{Department of Electronic Engineering \& State Key Lab of THz and Millimeter Waves, City University of Hong Kong, Kowloon, Hong Kong, China}
\affiliation{Department of Electronic Engineering, City University of Hong Kong, Kowloon, Hong Kong, China}
\author{Marko Lon{\^{c}}ar}\email{loncar@seas.harvard.edu}
\affiliation{John A. Paulson School of Engineering and Applied Sciences, Harvard University, Cambridge, Massachusetts 02138, USA}

\date{\today}
\title{Ultra-low loss integrated visible photonics using thin-film lithium niobate}

\begin{abstract}
Integrated photonics is a powerful platform that can improve the performance and stability of optical systems, while providing low-cost, small-footprint and scalable alternatives to implementations based on free-space optics. While great progress has been made on the development of low-loss integrated photonics platforms at telecom wavelengths, visible wavelength range has received less attention. Yet, many applications utilize visible or near-visible light, including those in optical imaging, optogenetics, and quantum science and technology. Here we demonstrate an ultra-low loss integrated visible photonics platform based on thin film lithium niobate on insulator. Our waveguides feature ultra-low propagation loss of 6 dB/m, while our microring resonators have an intrinsic quality factor of 11 million, both measured $\boldsymbol{\ }$at 637 nm wavelength. Additionally, we demonstrate an on-chip visible intensity modulator with an electro-optic bandwidth of 10 GHz, limited by the detector used. The ultra-low loss devices demonstrated in this work, together with the strong second- and third-order nonlinearities in lithium niobate, open up new opportunities for creating novel passive, and active devices for frequency metrology and quantum information processing in the visible spectrum range.
\end{abstract}

\maketitle

\section{1. INTRODUCTION}

Low-loss, active and integrated photonic platform operating at visible wavelengths is of great interest for applications ranging from quantum optics and metrology to bio-sensing and bio-medicine. For example, alkali and alkaline earth metals such as rubidium, cesium, calcium and sodium, the key elements for modern precision optical frequency metrology~[1--3], magnetometry~[4--6] and quantum computation ~[7--10], have their atomic transitions in visible and near-visible spectrum range.  In addition, integrated photonic circuits at visible wavelengths found their way into the fields such as optogenetics~[11,12] and bio-sensing~[13--15]. Furthermore, visible wavelength light is used for quantum state preparation~[16], manipulation and read out of color centers~[17], quantum dots~[18,19] and various quantum emitters in 2d materials~[20,21].   

Driven by these applications, several materials have been investigated as candidates for visible photonics platform, including SiO${}_{2\ }$~[22,23], Si${}_{3}$N${}_{4}$~[24--27], diamond ~[28--32], TiO${}_{2}$ ~[33] and AlN~[34]. With exception of AlN, all of these platforms are electro-optically passive and do not allow for fast control of optical signals. Here we show that lithium niobate (LN) is a promising integrated platform for visible photonics, owing to its wide transparency window (400 nm -- 5000 nm), and large electro-optic coefficient, $\mathrm{\sim}$30 times larger than that of AlN, and strong optical nonlinearity ~[35].  Our work builds on recently developed thin film lithium niobate (TFLN) substrates ~[36] and the novel fabrication method ~[37] which enabled realization of high-performance electro-optical (EO) modulators~[38--40] and Kerr  and EO frequency combs~[41,42] in telecom wavelength range (1500-1650nm). TFLN platform has also been used to demonstrate an effective generation of visible light via nonlinear processes such as second harmonic generation (SHG)~[43--47] and sum frequency generation (SFG)~[48].  In this work, we demonstrate low loss waveguides and Y-splitters, ultra-high-Q microring resonators, and electro optical (EO) modulator with 10GHz bandwidth (limited by the bandwidth of the detector used), operating at technologically relevant 600 -- 900 nm wavelength range.

\section{LOW-LOSS LN WAVEGUIDES AND HIGH-Q RESONATORS}

\noindent In our earlier work ~[37] focused on telecom LN devices, the main sources of waveguide loss were scattering, due to rough sidewalls, and linear absorption in SiO${}_{2}$ cladding.  The former is expected to be a lot more significant at visible wavelengths considered here, since Rayleigh scattering is proportional to ${\lambda }^{-4}$, where  $\lambda $  is the wavelength of light. Therefore, in order to minimize the interaction between the waveguide mode with sidewalls and oxide cladding we choose to work in rib configuration where the waveguide consist of a slab and a strip superimposed onto it (Fig. 1a). The waveguide parameters were chosen to satisfy three important conditions: i) single mode operation at wavelength of interest $\lambda =635$nm, for both transverse-electric (TE) and transverse-magnetic (TM) polarization; ii) minimal overlap between the optical mode, the waveguide sidewall and the oxide cladding; iii) bending loss  $\mathrm{<}$ 0.1 dB/cm for a bending radius of 50 $\mu m$. The latter was chosen in order to enable realization of compact high-Q ring resonators. Using numerical modeling (Lumerical) we found that these requirements are satisfied for the following waveguide parameters: strip height eT=180 nm, slab thickness wT=120 nm, and waveguide top width wW=480 nm. The sidewall angle was assumed to be 28 degrees with respect to vertical direction, and is the result of our fabrication process ~[37]. Fig. 1a and Fig. 1b show the mode profiles of this waveguide at 635 nm and 850 nm, respectively. Since one envisioned application of LN photonic platform is in non-linear multi-wavelength processes, we also evaluate the performance of our waveguide at telecom wavelengths (Fig. 1c). As expected, the optical mode is less confined at longer wavelength, which will result in larger optical losses. It should be noted that at elevated optical power,  additional nonlinear loss mechanisms may become relevant, including SHG, photorefractive effect ~[49] and thermal instability.

\begin{figure}
	\centering
	\includegraphics[angle=0,width=0.5\textwidth]{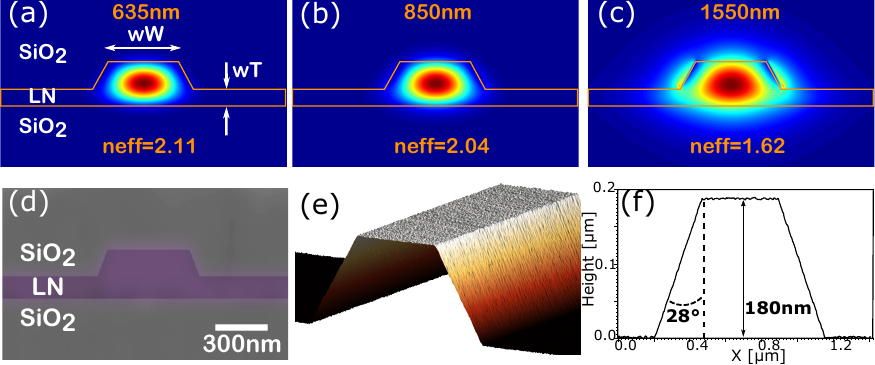}
	\caption{\label{fig1} \textbf{(a)-(c)} Finite element simulation of TE${}_{00}$ waveguide mode near three different wavelengths: 635 nm, 850 nm, and 1550 nm respectively. wW = 480 nm is waveguide width, wT = 120 nm is LN slab thickness \textbf{(d)} False-color SEM micrograph of the waveguide cross-section. \textbf{(e)} 2-D AFM scan on LN waveguide. \textbf{(f)} AFM line profile of LN waveguide.
}

\end{figure}

In order to characterize the optical losses, we fabricated microring resonators with various radii and coupling gaps (Fig. 2). The devices were fabricated on LN-on-insulator (LNOI) substrate (NANOLN) with 300 nm of X-cut LN layer on top of 2-$\mu$m thick thermally grown silicon dioxide layer. The structures were defined with electron beam lithography and the patterns were transferred through inductively coupled reactive ion etching with Ar+ plasma (ICP-RIE). Finally, the chip was cleaned and covered with 1-$\mu$m silicon dioxide, using plasma-enhanced chemical vapor deposition (PECVD). Finally, the waveguide facets were diced and polished.  The fabricated chips were inspected by scanning electron microscope (SEM) and atomic force microscope (AFM). Fig. 1d and Fig. 1e present a false-color SEM micrograph of a cladded device cross-section, and an AFM scan of a 500nm wide waveguide before cladding, respectively. On the sidewalls of the waveguides, the roughness, measured over 3 by 0.1-micron area, is found to be 0.7 nm RMS. The 28-degree sidewall angle was extracted from the 1d AFM profile (Fig.  1f).  

One of the main challenges in fabrication of microring resonators at visible wavelengths, is a narrow coupling gap needed for a single point coupling scheme. To overcome this difficulty, we implement a pulley coupling scheme where the coupling waveguides wraps around the ring. The exact coupling length was calculated at different wavelengths by using 3-D finite difference time domain (FDTD) simulations (Lumerical, FDTD). SEM micrographs of fabricated microring resonator and close-up zoom at coupling region are shown in Fig 2(a) and (b).  

\begin{figure}
	\centering
	\includegraphics[angle=0,width=0.5\textwidth]{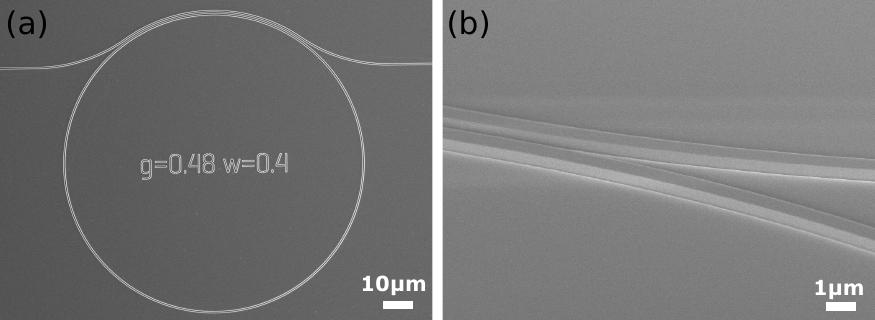}
	\caption{\label{fig2} \textbf{(a)} SEM micrograph of a fabricated microring resonator  (radius=100$\mu$m). \textbf{(b)} SEM image of coupling region. 
}
\end{figure}

The devices were characterized in 634-638 nm and 720 -- 850 nm range using New Focus Velocity and M2 SolsTis tunable lasers, respectively. Both lasers were calibrated using an external wavemeter and a home-built fiber-based Mach-Zehnder (MZ) interferometer. We launched a TE polarized laser beam into the coupling waveguide using a single mode visible lensed fiber (OZ Optics), and collected and detected transmitted light using another lensed fiber followed by a photodetector (New Focus,1801). The input polarization of the light was controlled by an external fiber-based polarization controller. To avoid the influence of the photorefractive effect and thermal instability, the devices were measured in low-power operation regime with tens of nanowatts of optical power (resulting in tens microwatts of circulating power inside the ring).   Typical fiber to chip coupling loss is $\mathrm{\sim}$ 6-10dB per facet due to mode mismatch from the fiber to the chip. 

\begin{figure}
	\centering
	\includegraphics[angle=0,width=0.5\textwidth]{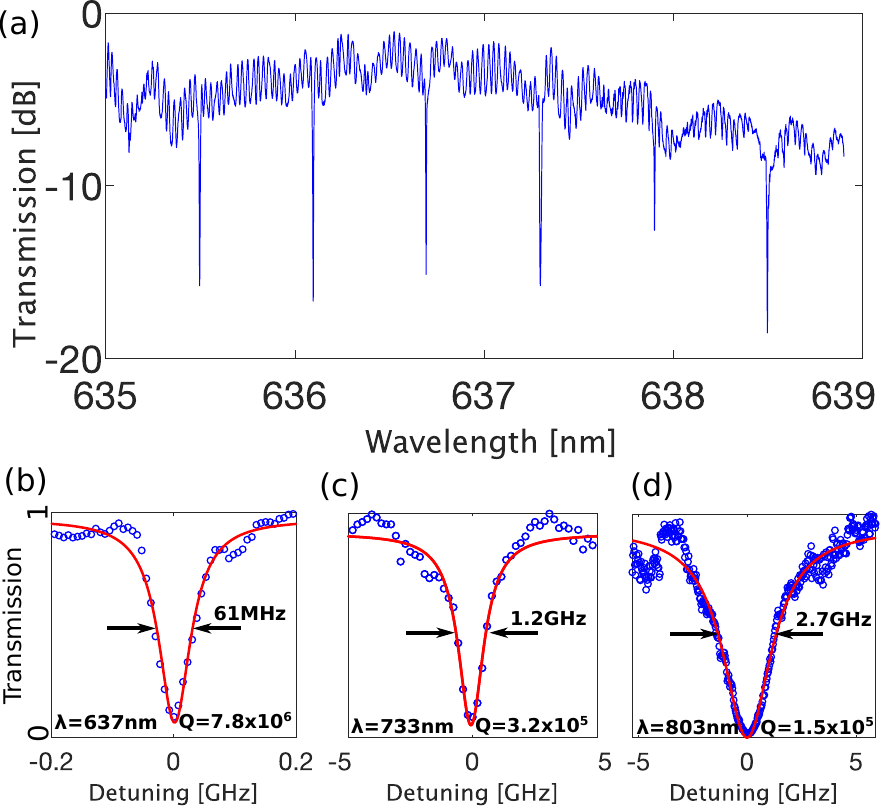}
	\caption{\label{fig3} \textbf{(a)} Measured transmission spectrum of TFLN microring cavity near 635nm wavelengths. \textbf{(b-d)} Fit of the resonance dips to Lorentzian function at wavelengths of 637nm, 730nm and 800nm respectively. Experimental data shown as blue dots and fit function shown as red line.}
\end{figure}

Fig. 3 shows the transmission spectra of a representative microring resonator measured at different wavelengths. By fitting the experimental results of under-coupled microring resonators with Lorentzian function, we estimate loaded quality factors ($Q_l\ $) of 7$.8\times {10}^6,\mathrm{3}.2\times {10}^5\mathrm{,and\ 1}.5\times {10}^5$ at wavelengths of 637 nm, 730 nm and 800 nm, respectively. These quality factors correspond to intrinsic quality factors of 1$.1\times {10}^7,\mathrm{5}.3\times {10}^5,\mathrm{2.7}\times {10}^5\ $ respectively.  We also characterized the same ring resonator at telecom wavelength range (1450-1650nm) using Santec 510 tunable laser (data not shown) and observed moderately high loaded Q factor of $\mathrm{1}.1\times {10}^5$ (intrinsic Q = $\mathrm{2}.3\times {10}^5$). As expected, quality factors decrease as wavelength increases due to reduced confinement of the optical mode, leading to increased overlap with waveguide sidewalls and cladding. Based on these results, we estimate the upper limit of a waveguide loss to be $\alpha \approx 6\ dB/m$ at wavelength 635 nm ~[50].  This value has same order of magnitude as previously reported loss value for TFLN waveguides at Infrared spectral range~[37]. It should be noted, that for TE polarized waveguide mode in the x-cut LN microring resonator, the refractive index will alternate between an ordinary and an extraordinary value of $n_o$ to $n_e$. However, for our microrings with large radii of more than 50 micron, such refractive index alternation happens in an adiabatic fashion, therefore does not impose any measurable additional optical loss to the system. To confirm this, the effect of polarization induced losses were analyzed by collecting the light at the output with a microscope objective and sending it through a polarizer. We do not observe any effects of TE/TM coupling or crosstalk. Importantly, our results show that LN ring resonators optimized for operation in red can support single mode low-loss operation across a wide wavelength range which is essential for envisioned applications in nonlinear optics, including second-harmonic generation, sum- and difference-frequency generation, and entangled photon pair generation. 

\section{ Y-SPLITTERS AND MACH-ZEHNDER INTERFEROMETR}

In addition to low loss waveguides and high-Q cavities, beam splitters and Mach-Zehnder interferometers (MZI) are key building blocks in integrated optics. There are many ways to realize an on-chip beam splitter such as MMI couplers~[51] Y-Junctions~[52] and directional couplers~[53].  Among these, Y-splitters are particularly interesting owing to their simplicity, tolerance to fabrication imperfections, and relatively wide bandwidth (hundreds of nanometers).  The main drawback of Y-splitters is their relatively large footprint of a few hundred microns.

\begin{figure}
	\centering
	\includegraphics[angle=0,width=0.5\textwidth]{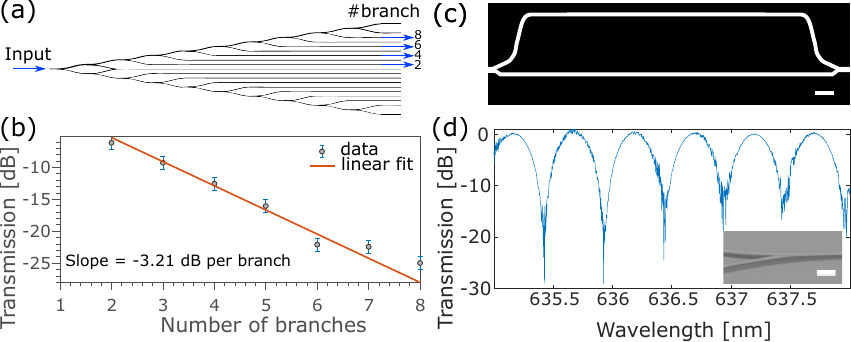}
	\caption{\label{fig4} \textbf{(a)} Mask layout of fabricated device.  \textbf{(b)} Measured transmission of cascaded Y-splitter tree as a function of number of Y-splitter branches. The orange line shows a linear fit with a slope of -3.21dB/splitter. \textbf{(c)} Dark field optical microscope image of the unbalanced MZI. Scale bar: 50 $\mu$m. \textbf{(d)} Measured transmission spectrum of the MZI showing extinction ratios of $\mathrm{\sim}$ 30 dB. Inset: SEM micrograph of Y-splitter section. Scale bar:  2 $\mu$m.
}
\end{figure}

To characterize visible TFLN beam splitters, we fabricated a ``Y-splitter tree`` (Fig. 4a). In this way, different output arms of the Y-splitter tree experience the same total waveguide length but different number of splitters. By comparing the transmission levels of different arms, the splitting ratios and the splitter losses can be extracted from a linear fit. Fig. 4(b) shows the normalized transmission of the cascaded Y-splitter tree, measured at 637 nm, as a function of number of Y-splitters in the cascade. Linear fit to experimental data shows a slope of -- 3.21 dB/splitter, indicating an excessive splitter loss of 0.21dB$\pm$0.01dB per Y-splitter.  

Fig. 4c shows a dark field optical image of a fabricated un-balanced MZI formed using two Y junctions and two low-loss waveguides.  Since top arm is longer, light propagating in it will accumulate additional phase compared to the light propagating in the bottom arm. After optical beams are recombined using a Y-splitter (Fig. 4d, inset), the difference in phase is converted into an amplitude modulation, resulting in constructive and destructive interference (Fig. 4c). An important figure of merit for MZI is the extinction ratio (ER), which is the ratio between the amplitude of constructively and destructively interfered light. In our devices the highest measured ER is $\mathrm{\sim}$ 30 dB, and is larger than 15 dB across the measured wavelength range.  Effects such as polarization mixing and higher order mode coupling in the Y-splitter are likely the cause of reduced extinction ratio at longer wavelengths. This can be improved by further optimization of Y-splitter design. 

\section{ INTENSITY MODULATOR }

\noindent Important advantage of LN visible photonic platform over competing platforms is the ability to realize efficient electro-optic modulators and optical switches and routers. To demonstrate this, we fabricated on-chip amplitude modulators which consist of unbalanced MZI with embedded active phase shifters in both interferometer's paths. Coplanar Ground-Signal-Ground (GSG) transmission line was used to deliver RF fields. The active phase shifters were fabricated in an additional lithography step followed by evaporation and lift-off of gold electrodes. The gap between the electrodes is 5 $\mu$m.  The optical microscope image of the fabricated structure is shown in Fig. 5a. To characterize the DC performance of the device we measured a normalized transmission of the device as a function of applied voltage. The voltage required for inducing a phase change of $\pi$ is called a half-wave voltage (\textit{${V}_{\pi }$}). We found a \textit{${V}_{\pi}$} of 8V for a 2 mm long device which translates into a voltage-length product (\textit{${V}_{\pi}L $}) of 1.6$Vcm$. This value is slightly better than previously reported infrared (IR) TFLN modulator ~[38] since the same refractive index shift would induce a larger phase accumulation at shorter wavelengths.

\begin{figure}
	\centering
	\includegraphics[angle=0,width=0.5\textwidth]{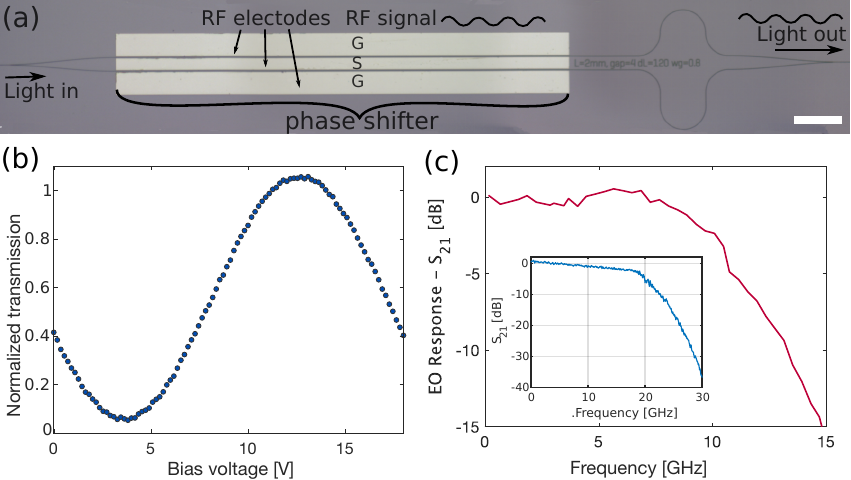}
	\caption{\label{fig5} \textbf{(a)} Optical image of the fabricated LN amplitude modulator. \textbf{(b)} Measured normalized transmission versus applied DC voltage showing a half-wave voltage of 8V for a 2 mm long device at a wavelength of 850nm. \textbf{(c)} Measured electro-optical response of the amplitude modulator. The 3dB cutoff frequency is $\mathrm{\sim}$ 10GHz limited by detector. Inset: Measured electrical insertion loss (\textit{S}${}_{21}$ parameters) shows an electrical (3dB) bandwidth of 17GHz.
}
\end{figure}

The electro-optic bandwidth of our modulator was measured using a vector network analyzer (Agilent E8364B). The optical signal from the modulator was sent to a high-speed avalanche photo diode (APD, EOT ET-4000A, bandwidth 10GHz). RF measurements were performed by using 50-$\Omega$, 40 GHz RF probes and all the results were normalized relative to RF cable losses. Fig. 5b shows the measured electro-optic response of our modulator at a wavelength of 850 nm. We measure our modulator with 100$\mu$W of optical power inside the device for several hours without observing any power instability or performance degradation in modulator operation. We measured the electro-optical 3-dB bandwidth to be 10 GHz and is currently limited by the bandwidth of high gain photo-detector used. To confirm this, we measured the electrical bandwidth (\textit{${S}_{21}$} parameter) of our coplanar transmission line (Fig.5c, inset) and found the electrical 3dB bandwidth to be 17 GHz. This value could be further increased by improving the design of the microwave coplanar transmission lines~[54]. We note that our numerical modeling indicates that modulation bandwidth is not limited by group velocity mismatch between the optical and RF signals. In the case of low RF propagation loss, the modulation bandwidth limit is inversely proportional to the product of waveguide length and the group index mismatch $\mathrm{\Delta}$\textit{${n}_{g}$}~[55].  In our case, 2mm long device with $\mathrm{\Delta}$\textit{${n}_{g}$ }=0.12, the bandwidth limit due to velocity mismatch is BW=300GHz and it is not limiting the EO performance of our modulator. We also note that the 10 GHz bandwidth is sufficient for many practical applications at visible spectrum range, including frequency-modulation spectroscopy~[56] or Pound-Drever-Hall laser locking technique~[57] .

\section{CONCLUSIONS}

In conclusion, we have demonstrated an ultra-low loss platform for integrated photonics at visible wavelengths, which achieved ultralow linear propagation losses. Additionally, we demonstrate an on-chip intensity modulator with electro-optic bandwidth of 10GHz and low voltage-length product of 1.6Vcm. We believe LN will become a powerful candidate for integrated on-chip photonic applications such as active light manipulation and wavelength conversion at visible wavelength range, as well as other applications such as combination with quantum emitters and alkali metals and will motivate future studies in the field of active photonic devices at visible wavelength range.

This work is supported in part by the National Science Foundation (NSF) (ECCS-1609549, ECCS-1740296 E2CDA), Defense Advanced Research Projects Agency (DARPA) (W31P4Q-15-1-0013) and City University of Hong Kong Start-up Funds.
We thank C. Reimer for feedback on the manuscript. Lithium niobate devices were fabricated in the Center for Nanoscale Systems (CNS) at Harvard, a member of the National Nanotechnology Infrastructure Network, supported by the NSF under award no. 1541959.

\section{REFERENCES}
\begin{enumerate}
\item J. C. Bergquist, "Frequency Standards and Metrology," in Frequency Standards and Metrology (WORLD SCIENTIFIC, 1996), pp. 1--574.
\item   T. Udem, R. Holzwarth, and T. W. H\"{a}nsch, "Optical frequency metrology," Nature 416, 233--237 (2002).
\item  J. Kitching, E. A. Donley, S. Knappe, M. Hummon, A. T. Dellis, J. Sherman, K. Srinivasan, V. A. Aksyuk, Q. Li, D. Westly, B. Roxworthy, and A. Lal, "NIST on a Chip: Realizing SI units with microfabricated alkali vapour cells," J. Phys. Conf. Ser. 723, 012056 (2016).
\item  L. W. Parsons and Z. M. Wiatr, "Rubidium vapour magnetometer," J. Sci. Instrum. 39, 292--300 (1962).

\item    H. Korth, K. Strohbehn, F. Tejada, A. G. Andreou, J. Kitching, S. Knappe, S. J. Lehtonen, S. M. London, and M. Kafel, "Miniature atomic scalar magnetometer for space based on the rubidium isotope 87Rb.," J. Geophys. Res. Sp. Phys. 121, 7870--7880 (2016).
\item    S. J. Smullin, I. M. Savukov, G. Vasilakis, R. K. Ghosh, and M. V. Romalis, "Low-noise high-density alkali-metal scalar magnetometer," Phys. Rev. A 80, 033420 (2009).
\item    A. I. Lvovsky, B. C. Sanders, and W. Tittel, "Optical quantum memory," Nat. Photonics 3, 706--714 (2009).
\item   K. F. Reim, P. Michelberger, K. C. Lee, J. Nunn, N. K. Langford, and I. A. Walmsley, "Single-Photon-Level Quantum Memory at Room Temperature," Phys. Rev. Lett. 107, 053603 (2011).
\item    O. Katz and O. Firstenberg, "Light storage for one second in room-temperature alkali vapor," Nat. Commun. 9, 2074 (2018).
\item    D. F. Phillips, A. Fleischhauer, A. Mair, R. L. Walsworth, and M. D. Lukin, "Storage of Light in Atomic Vapor," Phys. Rev. Lett. 86, 783--786 (2001).
\item    E. Shim, Y. Chen, S. Masmanidis, and M. Li, "Multisite silicon neural probes with integrated silicon nitride waveguidesand gratings for optogenetic applications," Sci. Rep. 6, 22693 (2016).
\item   E. Segev, J. Reimer, L. C. Moreaux, T. M. Fowler, D. Chi, W. D. Sacher, M. Lo, K. Deisseroth, A. S. Tolias, A. Faraon, and M. L. Roukes, "Patterned photostimulation via visible-wavelength photonic probes for deep brain optogenetics.," Neurophotonics 4, 011002 (2017).
\item .   P. Muellner, E. Melnik, G. Koppitsch, J. Kraft, F. Schrank, and R. Hainberger, "CMOS-compatible Si3N4 Waveguides for Optical Biosensing," Procedia Eng. 120, 578--581 (2015).
\item   T. Claes, W. Bogaerts, and P. Bienstman, "Experimental characterization of a silicon photonic biosensor consisting of two            cascaded ring resonators based on the Vernier-effect and introduction of a curve fitting            method for an improved detection limit," Opt. Express 18, 22747--22761 (2010).
\item  I. Goykhman, B. Desiatov, and U. Levy, "Ultrathin silicon nitride microring resonator for biophotonic applications at 970 nm wavelength," Appl. Phys. Lett. 97, 81108 (2010).
\item   I. Aharonovich, D. Englund, and M. Toth, "Solid-state single-photon emitters," Nat. Photonics 10, 631--641 (2016).
\item    I. Aharonovich and E. Neu, "Diamond Nanophotonics," Adv. Opt. Mater. 2, 911--928 (2014).
\item   Y. Chen, A. Ryou, M. R. Friedfeld, T. Fryett, J. Whitehead, B. M. Cossairt, and A. Majumdar, "Deterministic Positioning of Colloidal Quantum Dots on Silicon Nitride Nanobeam Cavities," Nano Lett. 18, 6404--6410 (2018).
\item    Y.-S. Park, S. Guo, N. S. Makarov, and V. I. Klimov, "Room Temperature Single-Photon Emission from Individual Perovskite Quantum Dots," ACS Nano 9, 10386--10393 (2015).
\item    Y.-M. He, G. Clark, J. R. Schaibley, Y. He, M.-C. Chen, Y.-J. Wei, X. Ding, Q. Zhang, W. Yao, X. Xu, C.-Y. Lu, and J.-W. Pan, "Single quantum emitters in monolayer semiconductors," Nat. Nanotechnol. 10, 497--502 (2015).
\item .   J. Kern, I. Niehues, P. Tonndorf, R. Schmidt, D. Wigger, R. Schneider, T. Stiehm, S. Michaelis de Vasconcellos, D. E. Reiter, T. Kuhn, and R. Bratschitsch, "Nanoscale Positioning of Single-Photon Emitters in Atomically Thin WSe 2," Adv. Mater. 28, 7101--7105 (2016).
\item    Y. Gong and J. Vu\^{c}kovi\'{c}, "Photonic crystal cavities in silicon dioxide," Appl. Phys. Lett. 96, 031107 (2010).
\item    S. H. Lee, D. Y. Oh, Q.-F. Yang, B. Shen, H. Wang, K. Y. Yang, Y.-H. Lai, X. Yi, X. Li, and K. Vahala, "Towards visible soliton microcomb generation," Nat. Commun. 8, 1295 (2017).
\item    P. Mu\~{n}oz, G. Mic\'{o}, L. A. Bru, D. Pastor, D. P\'{e}rez, J. D. Dom\'{e}nech, J. Fern\'{a}ndez, R. Ba\~{n}os, B. Gargallo, R. Alemany, A. M. S\'{a}nchez, J. M. Cirera, R. Mas, and C. Domí\textbf{}nguez, "Silicon Nitride Photonic Integration Platforms for Visible, Near-Infrared and Mid-Infrared Applications.," Sensors (Basel). 17, (2017).
\item    S. Romero-García, F. Merget, F. Zhong, H. Finkelstein, and J. Witzens, "Silicon nitride CMOS-compatible platform for integrated photonics applications at visible wavelengths," Opt. Express 21, 14036 (2013).
\item    E. Shah Hosseini, S. Yegnanarayanan, A. H. Atabaki, M. Soltani, and A. Adibi, "High quality planar silicon nitride microdisk resonators for integrated photonics in the visible wavelength range," Opt. Express 17, 14543 (2009).
\item    M. Khan, T. Babinec, M. W. McCutcheon, P. Deotare, and M. Lon{\^{c}}ar, "Fabrication and characterization of high-quality-factor silicon nitride nanobeam cavities," Opt. Lett. 36, 421 (2011).
\item    M. J. Burek, Y. Chu, M. S. Z. Liddy, P. Patel, J. Rochman, S. Meesala, W. Hong, Q. Quan, M. D. Lukin, and M. Lon{\^{c}}ar, "High quality-factor optical nanocavities in bulk single-crystal diamond," Nat. Commun. 5, 5718 (2014).
\item    P. Latawiec, V. Venkataraman, A. Shams-Ansari, M. Markham, and M. Lon{\^{c}}ar, "Integrated diamond Raman laser pumped in the near-visible," Opt. Lett. 43, 318 (2018).
\item   B. Sotillo, V. Bharadwaj, J. Hadden, S. Rampini, A. Chiappini, T. Fernandez, C. Armellini, A. Serpeng\"{u}zel, M. Ferrari, P. Barclay, R. Ramponi, S. Eaton, B. Sotillo, V. Bharadwaj, J. P. Hadden, S. Rampini, A. Chiappini, T. T. Fernandez, C. Armellini, A. Serpeng\"{u}zel, M. Ferrari, P. E. Barclay, R. Ramponi, and S. M. Eaton, "Visible to Infrared Diamond Photonics Enabled by Focused Femtosecond Laser Pulses," Micromachines 8, 60 (2017).
\item    L. Li, T. Schr\"{o}der, E. H. Chen, H. Bakhru, and D. Englund, "One-dimensional photonic crystal cavities in single-crystal diamond," Photonics Nanostructures - Fundam. Appl. 15, 130--136 (2015).
\item   P. Rath, S. Ummethala, C. Nebel, and W. H. P. Pernice, "Diamond as a material for monolithically integrated optical and optomechanical devices," Phys. status solidi 212, 2385--2399 (2015).
\item    J. T. Choy, J. D. B. Bradley, P. B. Deotare, I. B. Burgess, C. C. Evans, E. Mazur, and M. Lon{\^{c}}ar, "Integrated TiO\_2 resonators for visible photonics," Opt. Lett. 37, 539 (2012).
\item    C. Xiong, W. H. P. Pernice, X. Sun, C. Schuck, K. Y. Fong, and H. X. Tang, "Aluminum nitride as a new material for chip-scale optomechanics and nonlinear optics," New J. Phys. 14, 095014 (2012).
\item    R. S. Weis and T. K. Gaylord, "Lithium niobate: Summary of physical properties and crystal structure," Appl. Phys. A Solids Surfaces 37, 191--203 (1985).
\item P. Rabiei and P. Gunter, "Optical and electro-optical properties of submicrometer lithium niobate slab waveguides prepared by crystal ion slicing and wafer bonding," Appl. Phys. Lett. 85, 4603--4605 (2004).
\item    M. Zhang, C. Wang, R. Cheng, A. Shams-Ansari, and M. Lon{\^{c}}ar, "Monolithic ultra-high-Q lithium niobate microring resonator," Optica 4, 1536 (2017).
\item    C. Wang, M. Zhang, X. Chen, M. Bertrand, A. Shams-Ansari, S. Chandrasekhar, P. Winzer, and M. Lon{\^{c}}ar, "Integrated lithium niobate electro-optic modulators operating at CMOS-compatible voltages," Nature 562, 101--104 (2018).
\item    A. Rao, A. Patil, P. Rabiei, A. Honardoost, R. DeSalvo, A. Paolella, and S. Fathpour, "High-performance and linear thin-film lithium niobate Mach--Zehnder modulators on silicon up to 50\ \ GHz," Opt. Lett. 41, 5700 (2016).
\item    A. J. Mercante, S. Shi, P. Yao, L. Xie, R. M. Weikle, and D. W. Prather, "Thin film lithium niobate electro-optic modulator with terahertz operating bandwidth," Opt. Express 26, 14810 (2018).
\item    C. Wang, M. Zhang, R. Zhu, H. Hu, and M. Lon{\^{c}}ar, "Monolithic photonic circuits for Kerr frequency comb generation, filtering and modulation," (2018).
\item M. Zhang, B. Buscaino, C. Wang, A. Shams-Ansari, C. Reimer, R. Zhu, J. Kahn, and M. Lon{\^{c}}ar, "Broadband electro-optic frequency comb generation in an integrated microring resonator," (2018).
\item    C. Wang, X. Xiong, N. Andrade, V. Venkataraman, X.-F. Ren, G.-C. Guo, and M. Lon{\^{c}}ar, "Second harmonic generation in nano-structured thin-film lithium niobate waveguides," Opt. Express 25, 6963 (2017).
\item    R. Luo, Y. He, H. Liang, M. Li, and Q. Lin, "Highly tunable efficient second-harmonic generation in a lithium niobate nanophotonic waveguide," Optica 5, 1006 (2018).
\item    C. Wang, C. Langrock, A. Marandi, M. Jankowski, M. Zhang, B. Desiatov, M. M. Fejer, and M. Lon{\^{c}}ar, "Ultrahigh-efficiency wavelength conversion in nanophotonic periodically poled lithium niobate waveguides," Optica 5, 1438 (2018).
\item    R. Geiss, S. Saravi, A. Sergeyev, S. Diziain, F. Setzpfandt, F. Schrempel, R. Grange, E.-B. Kley, A. T\"{u}nnermann, and T. Pertsch, "Fabrication of nanoscale lithium niobate waveguides for second-harmonic generation," Opt. Lett. 40, 2715 (2015).
\item    R. Luo, Y. He, H. Liang, M. Li, and Q. Lin, "Semi-nonlinear nanophotonic waveguides for highly efficient second-harmonic generation," (2018).
\item   Z. Hao, J. Wang, S. Ma, W. Mao, F. Bo, F. Gao, G. Zhang, and J. Xu, "Sum-frequency generation in on-chip lithium niobate microdisk resonators," Photonics Res. 5, 623 (2017).
\item    A. A. Savchenkov, A. B. Matsko, D. Strekalov, V. S. Ilchenko, and L. Maleki, "Enhancement of photorefraction in whispering gallery mode resonators," Phys. Rev. B 74, 245119 (2006).
\item C. Zhang, L. R. Dalton, P. Rabiei, and W. H. Steier, "Polymer Micro-Ring Filters and Modulators," J. Light. Technol. Vol. 20, Issue 11, pp. 1968- 20, 1968 (2002).
\item Z. Wang, Z. Fan, J. Xia, S. Chen, and J. Yu, "1$\mathrm{\times}$8 Cascaded Multimode Interference Splitter in Silicon-On-Insulator," Jpn. J. Appl. Phys. 43, 5085--5087 (2004).
\item S. H. Tao, Q. Fang, J. F. Song, M. B. Yu, G. Q. Lo, and D. L. Kwong, "Cascade wide-angle Y-junction 1 $\mathrm{\times}$ 16 optical power splitter based on silicon wire waveguides on silicon-on-insulator," Opt. Express 16, 21456 (2008)
\item  H. Yamada, Tao Chu, S. Ishida, and Y. Arakawa, "Optical directional coupler based on Si-wire waveguides," IEEE Photonics Technol. Lett. 17, 585--587 (2005).
\item    Xiang Zhang and T. Miyoshi, "Optimum design of coplanar waveguide for LiNbO/sub 3/ optical modulator," IEEE Trans. Microw. Theory Tech. 43, 523--528 (1995).
\item    D. Janner, M. Belmonte, and V. Pruneri, "Tailoring the Electrooptic Response and Improving the Performance of Integrated LiNbO3 Modulators by Domain Engineering," J. Light. Technol. Vol. 25, Issue 9, pp. 2402-2409 25, 2402--2409 (2007).
\item    G. C. Bjorklund, M. D. Levenson, W. Lenth, and C. Ortiz, "Frequency modulation (FM) spectroscopy," Appl. Phys. B Photophysics Laser Chem. 32, 145--152 (1983).
\item    R. W. P. Drever, J. L. Hall, F. V. Kowalski, J. Hough, G. M. Ford, A. J. Munley, and H. Ward, "Laser phase and frequency stabilization using an optical resonator," Appl. Phys. B Photophysics Laser Chem. 31, 97--105 (1983).

\end{enumerate}

\end{document}